\documentclass[conference]{IEEEtran}
\usepackage[fontsize=9.9]{scrextend}

\usepackage[utf8x]{inputenc}
\usepackage[table,xcdraw]{xcolor}
\usepackage[pdfa]{hyperref}
\usepackage{balance}
\usepackage{cite}
\usepackage{csquotes}
\usepackage[english]{babel}
\usepackage{amsmath,amssymb,amsfonts}
\usepackage{algorithmic}
\usepackage{graphicx}
\usepackage{textcomp}
\usepackage{xcolor}
\usepackage{jabbrv}
\usepackage[inline,shortlabels]{enumitem}

\setlist[itemize]{noitemsep, topsep=0pt}
\setlist[enumerate]{noitemsep, topsep=0pt}
\def\mysize{9pt}
\DefineJournalAbbreviation{Proceedings}{Proc}
\DefineJournalAbbreviation{Neural}{Neur}
\DefineJournalAbbreviation{Advances}{Adv}
\DefineJournalAbbreviation{Information}{Inf}
\DefineJournalAbbreviation{Processing}{Process}

\title{Acoustics-specific Piano Velocity Estimation}

\author{
	\IEEEauthorblockN{Federico Simonetta, Stavros Ntalampiras, Federico
		Avanzini}
	\IEEEauthorblockA{
		LIM --- Music Informatics Laboratory\\
		Department of Computer Science\\
		University of Milan\\
		Email: \{name.surname\}@unimi.it    }
}

\begin{document}
\maketitle

\begin{abstract}

  Motivated by the state-of-art psychological research, we note that a piano
  performance transcribed with existing Automatic Music Transcription (AMT)
  methods cannot be successfully resynthesized without affecting the artistic
  content of the performance. This is due to 1) the different mappings between
  MIDI parameters used by different instruments, and 2) the fact that musicians
  adapt their way of playing to the surrounding acoustic environment. To face
  this issue, we propose a methodology to build acoustics-specific AMT systems
  that are able to model the adaptations that musicians apply to convey their
  interpretation. Specifically, we train models tailored for virtual instruments
  in a modular architecture that takes as input an audio recording and the
  relative aligned music score, and outputs the acoustics-specific velocities of
  each note. We test different model shapes and show that the proposed
  methodology generally outperforms the usual AMT pipeline which does not
  consider specificities of the instrument and of the acoustic environment.
  Interestingly, such a methodology is extensible in a straightforward way since
  only slight efforts are required to train models for the inference of other
  piano parameters, such as pedaling.

\end{abstract}


\section{Introduction}

Automatic Music Transcription (AMT) consists in the analysis of music audio
recordings to discover semantically meaningful events, such as notes,
instruments and chords. In this work, we  refer to those AMT methods that
take as input an audio recording and output MIDI-like representations of the
music performance. Two main methodologies for AMT exist, i.e. Non-negative
Matrix Factorization (NMF) and Deep Learning (DL) (for a thorough review
see~\cite{benetos2019automatic}). During the last 4 years, DL has tremendously
advanced the state-of-art of AMT, especially for piano music
recordings~\cite{hawthorne2018onsets,kong2021highresolution,yan2021skipping}.
Nonetheless, no attention has been placed on the influence of the instrument and environment
acoustics on the AMT output.

\begin{figure}
	\begin{center}
		\includegraphics[width=0.4\textwidth, trim= 0 40 0 0]{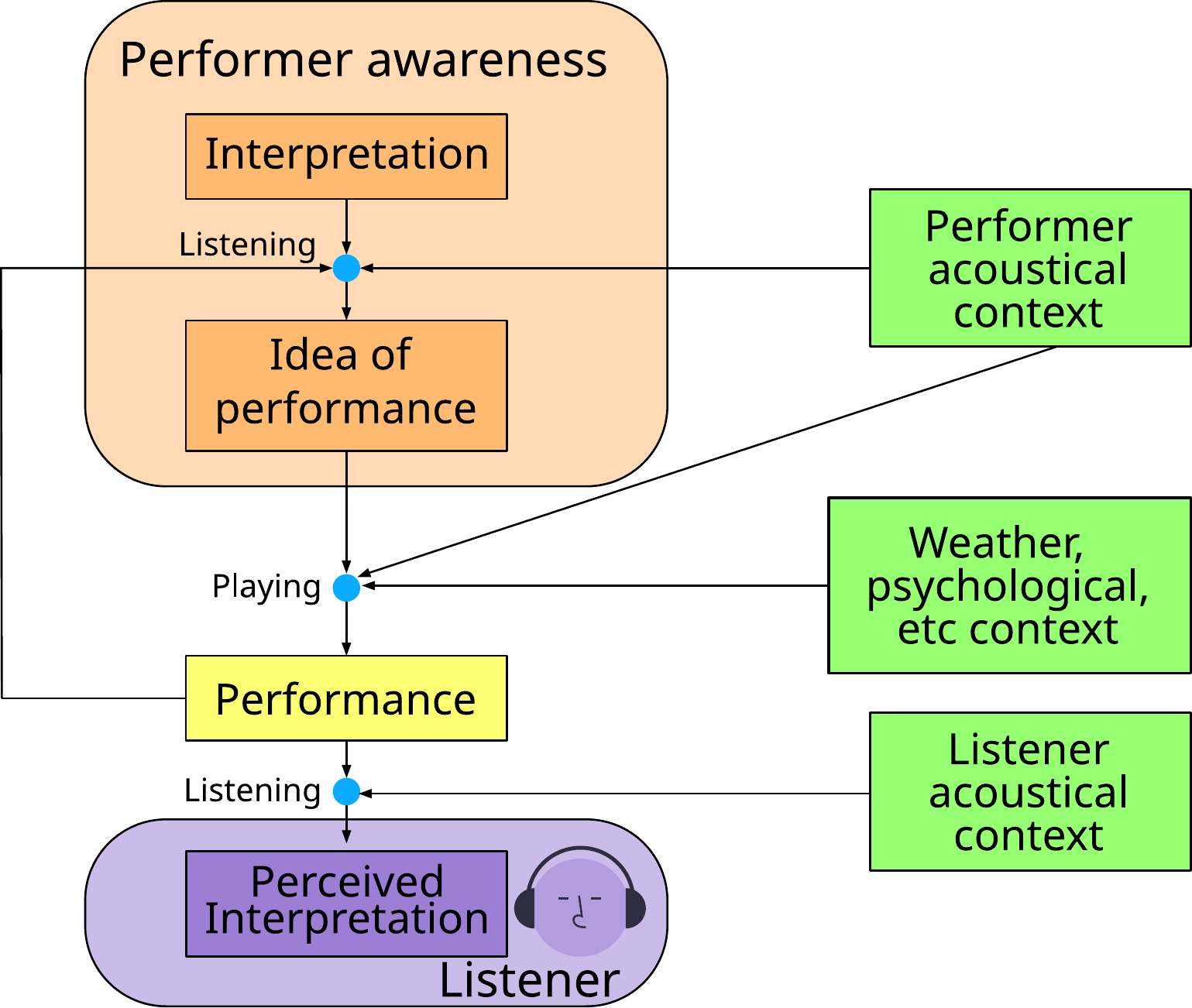}
	\end{center}
	\caption{Diagram of the influence of the acoustic context on the performance
		as outlined in the related literature. First, the performer creates an
		interior representation of the performance that it wants to achieve
		%-- what we name 
		(the ``interpretation''). Then, based on the acoustic environment, the
		musician consciously adapts its idea of performance. Successively, based on
		the feedback from the acoustic environment, it adjusts the performance in
		a circular feedback. Other non-acoustic factors
		%may influence this
		%phenomenon in an unconscious way, but the related literature provides no
		%evidence of them; nevertheless, they
		are included in the diagram to
		increase its generality. Finally, the listener itself understands the
		artistic intention based on the its own acoustic context.}

	\label{fig:contexts}
\end{figure}

Various evidence exist proving such influence~\cite{kob2020effect}.
%of the environment acoustics on the musician
Different research groups focused their efforts in the understanding
of why and how music performers adapt their way of playing to the surrounding
acoustics. Analyzed instruments include strings, wind instruments, and piano,
while the major finding is that the adaptations applied by musicians influence
the timbre, the amplitude dynamics, and the timing. Overall, psychologists agree
in the identification of an ``interior idea of the performance'' that the
musician adapts to the surrounding  acoustic environment. A diagram of the
phenomenon is shown in Fig.~\ref{fig:contexts} while an an overview of existing works and methodologies is
presented in~\cite{kob2020effect}.
However, all the existing studies are directed towards the understanding of the
factors characterizing room acoustics. Conversely, they rarely consider the
listener's perception and never take into account indirect factors that can
effectively change the acoustics of the instrument, such as  temperature and
humidity.

%trim=left bottom right top

The long-term use-case scenario that motivates the present work is an audio
resynthesis application that allows musicians to achieve studio-quality music
tracks by recording with cheap microphones and resynthesizing with high-quality
virtual instruments, regardless of the acoustic environment of
the source recording. The challenge is in the extraction of the artistic
elements that constitute the expressive intention of the performance and in
their re-rendering in a new known acoustic context -- e.g. a virtual
instrument. To this aim, in a previous work~\cite{simonetta2022perceptual}, we
%tried to use
resynthesized the output of a state-of-art AMT model and the ground-truth MIDI
in a new acoustic context, such as a high-quality piano
virtual instrument. We found that the perceived artistic content
of the original and of the new acoustic context was different for both the AMT model and
the ground-truth MIDI data recorded on a sensorized-piano.

As in our previous work~\cite{simonetta2022perceptual}, we use a
terminological distinction to identify two different levels of the performance.
Namely, the term ``performance'' includes the set of physical events that constitute
the act of playing; on the contrary, ``interpretation'' comprises the ideal
performance that the musician wants to convey and that it adapts to the
surrounding acoustic context -- see Fig.~\ref{fig:contexts}. As it emerges from the current discussion, the
performance depends on the acoustic environment where it happens, while the
interpretation is independent from it. Such a distinction is discussed more in
depth and studied from the perceptual perspective
in~\cite{simonetta2022perceptual}.

The contribution of the present paper is  the development of acoustics-specific
models that are able to transfer the interpretation to a certain acoustic
context by modeling the adaptations
that musicians apply to convey their interior idea of performance. One previous
work faced a similar problem, but tried to simulate the adaptation phenomenon
applied by performers while knowing both the original and target
context~\cite{xu201905transferring}. The proposed method, instead, is
independent from the input context and can theoretically work using a recording
taken in any acoustic context as input.

Towards more complete systems, the proposed approach focuses on the influence of
the acoustic context on the piano velocity. The velocity is a
MIDI parameter that controls the intensity with which a piano key is pressed. In
the MIDI standard, the velocity is an integer value in $[0, 127]$, but there is
no agreement on the mapping from this values to the physics of the
instrument~\cite{dannenberg2006interpretation}.  Consequently, the same velocity
value does not sound the same in different virtual instruments nor
actuated/sensorized pianos. However, the same applies to real-world pianos and rooms, because
different piano/room combinations have different responses to the
same physical stimulus. This is  why musicians need to adapt
their performance to different acoustic environments.

The proposed system aims at 1) transcribing MIDI data that successfully convey
the interpretation content considering the target acoustic context, and 2)
modeling the adaptations to the velocity that musicians put in place in
different acoustic environments. We train a model to first extract the
interpretation and then adapt it to each specific acoustic setting. The long-term goal
is  a system that can transcribe any piano recording in a way that can
be resynthesized with one of the acoustic contexts seen during training.

We perform thorough tests with different DL
models and demonstrate that acoustics-specifics strategies generally outperform the traditional non acoustics-specifics AMT systems. Targeting at full reproducibility of the
conducted experiments, the implementation is available
online\footnote{\label{foot:github}\url{https://limunimi.github.io/MIA/}}.

\section{Experiment Overview}
\label{sec:experimental_overview}

One of the main datasets available for piano AMT is the Maestro
dataset~\cite{hawthorne2018onsets}, containing 1276 audio recordings performed
on a Yamaha Disklavier, a highly precise sensorized piano. The performances were
recorded and for each audio, a precise MIDI recording is available. The whole
datasets provides 7.04 millions of notes and 198.7 hours of music.

Instead of collecting a new ad-hoc dataset, we leverage existing datasets
for piano AMT by resynthesizing various performances in multiple artificial
acoustic contexts. In this setting, a certain MIDI sequence rendered in two different contexts without any adaptation generates two outcomes with
different perceivable interpretations~\cite{simonetta2022perceptual}. Thus, the
same MIDI notes synthesized in different contexts without adaptation
should be considered as generated from different interpretations, even if they
have the same underlying MIDI data. Following this idea, we used the MIDI
ground-truth data to resynthesize the Maestro dataset~\cite{hawthorne2018onsets}
in manifold artificial contexts.

We designed an ad-hoc AMT model for estimating the performance parameters.
Since our aim is to understand whether the performance transcription error can
be reduced by considering the acoustic context, we focus on performance
parameter estimation in a controlled setting where note timings are known. In
this work, we limit our analysis to velocity estimation, but the proposed method
can be extended to %the estimation of 
new parameters.

In realizing an extensible AMT method, we use the perfectly aligned MIDI files
recorded by the Disklavier which are available in the Maestro dataset to inform
the transcription process. In a real-world scenario, a precise alignment of a
score can be obtained using the Audio-to-score method presented in the last year
MMSP conference~\cite{simonetta2021audiotoscore}. We applied Non-negative Matrix
Factorization (NMF)~\cite{lee2001algorithms} to perform a source-separation of
each single key of the piano and then analyzed the spectrogram representation of
each source-separated note. Namely, we compute the MFCCs
characterizing timbral aspects that are connected to the note velocity due to
the piano acoustics~\cite{bernays2014investigating} and are independent of the
non-linear amplitude distortions of the
microphones~\cite{jeong2020noteintensity,alias2016review}. Then, we employ a
Convolutional Neural Network (CNN) model to infer the velocity of each note.

The CNN model is split in two parts as follows:

\begin{enumerate}

	\item the ``encoder'', which  infers the interpretation by taking as input the
	      note-separated spectrogram with size $13\times30$ and one channel, where
	      $13$ is the number of MFCCs and $30$ is the number of considered time
	      frames;

	\item the ``performer'', which adapts the interpretation to the target
	      acoustic context by taking the latent output of the encoder and computing the
	      velocity in the target context.

\end{enumerate}

Let us observe a typical workflow for two MIDI files and two acoustic contexts:
\begin{enumerate}

	\item the two MIDI files are synthesized using respectively two acoustic
	      settings $X_0$ and $X_1$,  generating two audio waveforms $A_0$ and $A_1$;

	\item all the notes are extracted from $A_0$ and $A_1$ using the NMF
	      algorithm;

	\item each note spectrogram $n_{0i}$ from $A_0$ is processed by the encoder
	      producing a latent representation $L_{0i}$; that is also true for each note
	      $n_{1i}$  from $A_1$, that generates a latent representation $L_{1i}$;
	      if the underlying MIDI velocity for $n_{0k}$
	      and $n_{1h}$ was the same, their spectrogram will be different, as well as
	      their interpretation data, that we model with the latent representation
	      $L_{0k}$ and $L_{1h}$; this is mainly because the resynthesis did not adapt the
	      velocity to $X_0$ and $X_1$;

	\item each latent representation is then processed by the performer, which
	      infers the original velocity value, i.e.\ the same value for $n_{0k}$ and $n_{1h}$.

\end{enumerate}

Furthermore, we considered various strategies for acoustics-specificity. More specifically:

\begin{enumerate}

	\item the performer can be the same for every context, i.e.
	      context-independent, or it can be context-specific, meaning that in the
	      model there is a different performer for each context;

	\item since $L_{0k} \neq L_{1h}$, the latent space found by the encoder should  still include
	      the information required for identifying the input context; as such, it is
	      interesting to understand if enforcing this property facilitates the
	      learning process.

\end{enumerate}

The implementation of the first variable consists in building one performer per
context or one performer for every context. As regards to the second variable,
we add an additional branch to the model to classify the target context based on
the encoder output; this branch works as a learnable loss function that
separates the latent space according to the input context.

We compared the 4 different strategies resulting by the combinations of the
above-mentioned 2 boolean variables, i.e.

\begin{itemize}

	\item \textbf{Single-w/o}: one single performer and no context classification
	      -- this case corresponds to non acoustics-specific AMT;

	\item \textbf{Multiple-w/o}: one performer per context without using a context
	      classifier;

	\item \textbf{Single-with}: one single performer with context classifier;

	\item \textbf{Multiple-with}: one performer per context and context classification.

\end{itemize}

Since we aim at showing that knowing the target context is beneficial regardless
the approximation error, we repeat the experiments with various types of
function. More precisely, we define a set of hyper-parameters that determine the
shape of the neural network model and then perform a grid-search to explore how
the error changes when different model structures are used for the estimation.

All four strategies were tested for each point in the hyper-parameter space,
resulting in a highly computationally demanding experiment. We thoroughly tested 36
different model shapes, each with 4 different training strategies,
summing up to 144 trained models.

\section{Dataset}

This section describes the data creation process facilitating the proposed experiment.

\subsection{Resynthesis}
\label{sec:ch_formal_resynthesis}

We
designed an experiment based on the resynthesis of existing datasets. To this
end, we  developed
\texttt{pycarla}\footnote{\url{https://web.archive.org/web/20211213132835/https://pypi.org/project/pycarla/}},
a Python module that leverages the excellent
Carla\footnote{\url{https://web.archive.org/web/20211205195725/https://kx.studio/Applications:Carla}}
plugin host to synthesize MIDI messages both in real-time and offline using the
major audio plugin formats -- such as VST, AU, LV2, LADSPA, DSSI, SF2, SFZ.
%Even if \texttt{pycarla} has been designed to support long-running processes, such as the resynthesis of large music datasets, it is based on a complex pipeline, starting with the Jack\footnote{\url{https://web.archive.org/web/20211211061956/https://jackaudio.org/}} audio server, passing through Carla and ending with selected audio plugins. In this pipeline, various bugs, crashes, and errors may occur negatively affecting the synthesized audio. For this reason, various checks on the resynthesized audio have been carried out to identify potential synthesis errors.

\begin{table}[t]
	\centering
	\caption{Summary of the main characteristics of the 6 presets used for
		resynthesizing the Maestro~\cite{hawthorne2019enabling} dataset.}
	\begin{tabular}{|c|c|c|c|}
		\hline
		\textbf{ID} & \textbf{Velocity Map} & \textbf{Reverb} & \textbf{Instrument} \\ \hline
		0           & Linear                & Jazz Studio     & Steinway B Prelude  \\ \hline
		1           & Logarithmic           & Jazz Studio     & Steinway B Prelude  \\ \hline
		2           & Logarithmic           & Cathedral       & Steinway B Prelude  \\ \hline
		3           & Linear                & Jazz Studio     & Grotrian Cabaret    \\ \hline
		4           & Logarithmic           & Jazz Studio     & Grotrian Cabaret    \\ \hline
		5           & Logarithmic           & Cathedral       & Grotrian Cabaret    \\ \hline
	\end{tabular}
	\label{tab:ch_formal_presets}
\end{table}

We used 6 different presets for the physically modeled
virtual piano by Pianoteq, kindly provided for research purposes by
Modartt\footnote{\url{https://web.archive.org/web/20211112075858/https://www.modartt.com/}}.
Table~\ref{tab:ch_formal_presets} summarize the main characteristics of
each preset.

\subsection{Clustering}

The source dataset was Maestro~\cite{hawthorne2019enabling} and was used as
provided by ASMD library~\cite{simonetta2020automatic}. The Maestro dataset was
selected using the ASMD Python API; then, \emph{train}, \emph{validation}, and
\emph{test} splits were partitioned in 6 different subsets, each associated to
one of the presets in Table~\ref{tab:ch_formal_presets}, sampled so that they
were equally distributed across the dataset.
Similarly to stratified sampling, 6 sets were generated by unifying
subsets associated to the same preset, so that each set was still split in
\emph{train}, \emph{validation}, and \emph{test} sets. Each generated set was
resynthesized and saved to a new ASMD definition file.

Since each set consisted of only $1/6$
of the whole dataset, a uniform random sampling would have not grasped the underlying
distribution. Thus, we developed an
ad-hoc method to obtain equal-sized subsets maintaining a similar underlying
distribution as the parent set. To this aim, we extracted features from the MIDI
data and clustered them.

The 6 new subsets in each split were chosen as follows. First, one split at a
time among the already defined \emph{train}, \emph{validation}, and \emph{test}
was selected. Supposing that the chosen split has cardinality $K$, $C$ clusters
were created with $C = \lfloor K/6 + 1\rfloor$ and a target cardinality $t=6$
was set. Then, a redistribution policy is applied to the points of the clusters:
for each cluster with cardinality $< t$ -- a ``poor'' cluster -- we look for the
point nearest to that cluster's centroid and belonging to a cluster with
cardinality $> t$ -- a rich cluster -- and move that point to the poorer
cluster. The redistribution stops when all clusters have cardinality $\geq t$.
Since the redistribution algorithm moves points from rich clusters to the poor
ones, we named it ``Robin Hood'' redistribution policy. Having obtained $C$
clusters each with 6 samples, we partitioned the chosen split in 6 subsets as
follows:

\begin{enumerate}

	\item we randomized the order of subsets and clusters

	\item \label{item_first} we selected one point from each cluster using a
	      random uniform distribution and assigned it to one of the 6 subsets

	\item we did the same for the other 5 subsets

	\item we restarted from point~\ref{item_first} until every point is assigned to a cluster.

\end{enumerate}

Details about the clustering procedures and the features selected are provided
in the accompanying website.

\section{Note-separation}
\label{sec:ch_formal_separation}

\begin{figure}
	\center
	\includegraphics[width=0.5\textwidth]{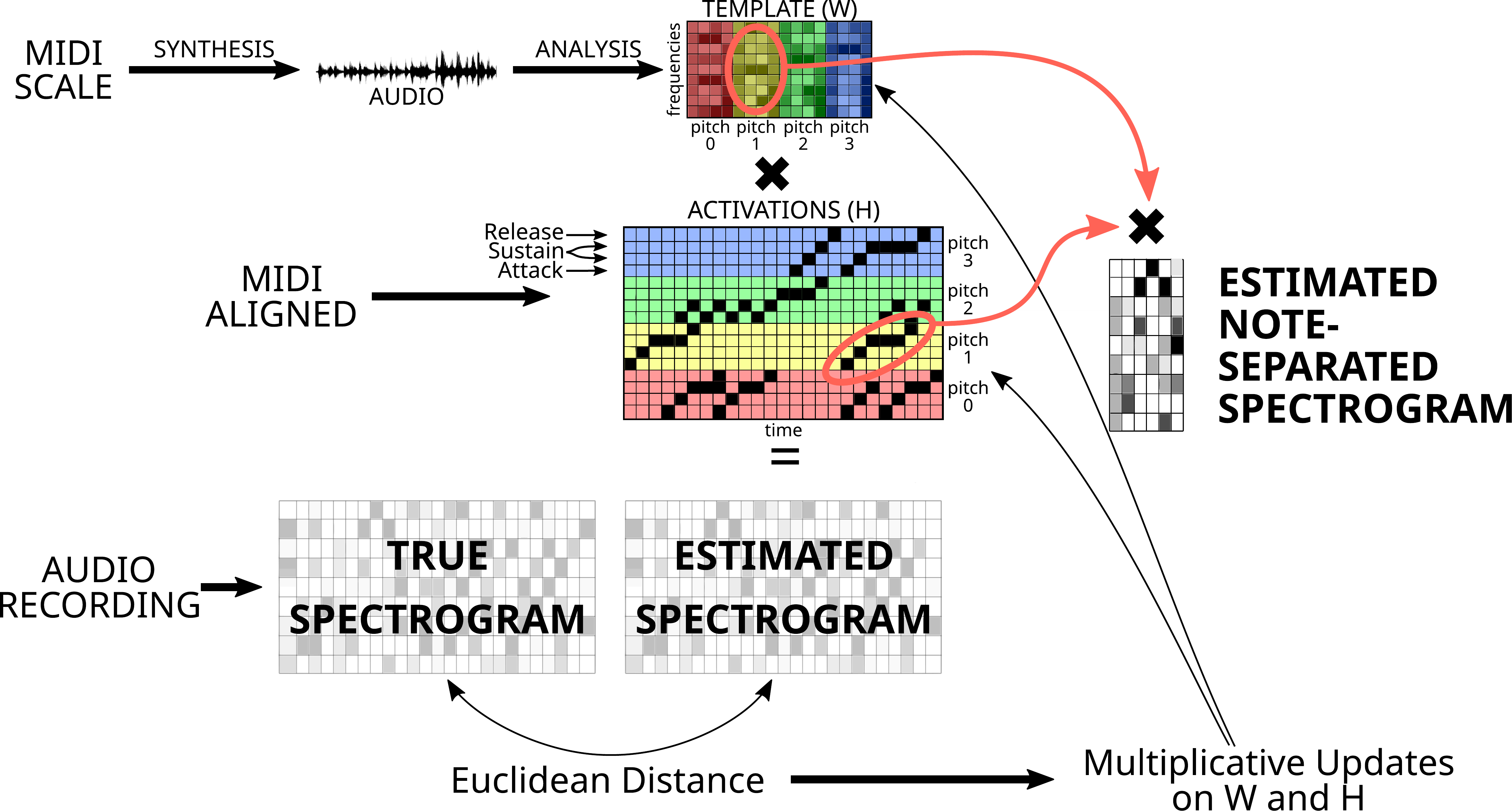}
	\caption{
		The entire NMF workflow. First, the initial template and activation matrices
		are computed. Then, the Euclidean distance between the estimated and the
		true spectrograms is used to multiplicatively update both the template and
		the activation matrix. Finally, only the relevant part of the template and
		activations are used for estimating the note-separated spectrogram. For simplicity, 4 columns are used to represent each note while during the experiments, we considered 30.
	}
	\label{fig:ch_formal_nmf}
\end{figure}

NMF has largely been used for score-informed
AMT~\cite{benetos2012scoreinformed,wang201710identifying,jeong2020noteintensity,simonetta2022perceptual}
and our application is mainly based on the existing literature. Using NMF, a
target non-negative matrix $S$ can be approximated with the multiplication
between a non-negative template matrix $W$ and a non-negative activation matrix
$H$. When applied to audio, $S$ is usually a time-frequency representation of
the audio recording, $W$ is the template matrix representing each audio source,
and $H$ represents the instants in which each source is active. As such, the
rows of $W$ represent frequency bins, the columns of $W$ and the rows of $H$
refer to sound sources, and the columns of $H$ are time frames. The $W$ and $H$
matrices are first initialized with some initial values and then updated until
some loss function comparing $S$ and $W \times H$ is minimized. Similarly to previous
works~\cite{simonetta2022perceptual,jeong2020noteintensity}, we build an initial
template matrix by analyzing a MIDI file synthesized with a further virtual
instrument and containing all the
88 piano keys playing with different velocities and duration. We initialize the
activation matrix using pianoroll-based information from the aligned MIDI file and proceed to
minimize the euclidean distance. The proposed
method for NMF is shown in figure~\ref{fig:ch_formal_nmf} and more
details are provided in the accompanying website.
Once the NMF algorithm is finished, we use the original perfectly aligned
activation matrix to select the region of a note in $H$ and $W$ to obtain its
approximated spectrogram separated from the rest of the recording. We  consider
the first 30 frames (690 ms) of each note, padding with 0 if the note is
shorter. We finally compute the first 13 MFCC features in each column of the
spectrogram using Essentia.

\section{Neural Network Models}
\label{sec:ch_formal_nn}

\begin{figure}
	\center
	\includegraphics[width=0.4\textwidth]{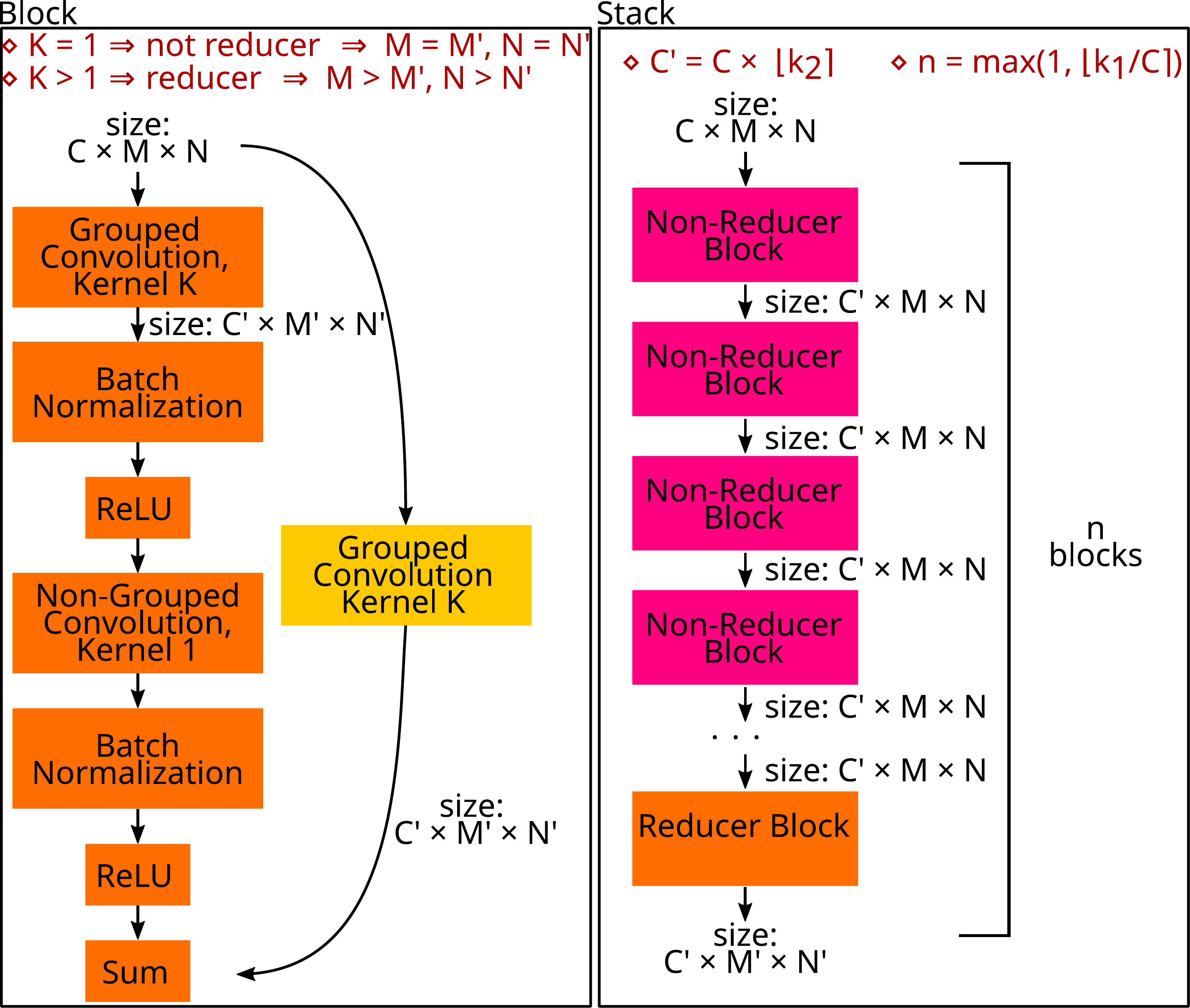}
	\caption{
		Architecture of the Residual blocks and stacks networks used
		in this work. $\lfloor \cdot \rceil$ denotes the rounding operation.
	}
	\label{fig:ch_formal_nn_model}
\end{figure}

For every function estimation, we use Convolutional Neural Networks (CNN)
with skip connections similarly to ResNet~\cite{he2016residual}. A schematic
representation of the proposed model building blocks is shown in
Figure~\ref{fig:ch_formal_nn_model}.

In ResNet, a building block is defined so that the output can have the same size
as the input (``not reducer'' block) or can be reduced (``reducer'' block); in
both cases, the output of each block is summed to the input to prevent the
vanishing gradients phenomenon and other degradation problems connected with the
increase of the network complexity~\cite{he2016residual}. Since they can
maintain the output size equal to the input, a virtually infinite number of
blocks can be put one after the other, and multiple stacks of blocks can be
concatenated to create arbitrarily large and complex networks without depending
on the input size.

In the proposed model, each block consists of the following elements:

\begin{itemize}

	\item a grouped convolutional layer
	      with kernel size $K$; if the block is a not-reducer, a padding is used;

	\item a batch-normalization layer;

	\item a ReLU non-linear activation;

	\item a non-grouped convolutional layer with kernel size 1 -- corresponding to
	      a linear combination of each data entry across channels;

	\item another batch-normalization layer;

	\item a final ReLU activation.

\end{itemize}

Furthermore, each block sums its output to the input processed with a grouped
convolutional layer having kernel size 1 if the block is a not reducer and  $K$
otherwise. Figure~\ref{fig:ch_formal_nn_model} better depicts the building of a
single block.

Multiple blocks can be put one after the other forming a stack. In each stack,
the first block changes the number of channels, while the rest keeps
it constant. Moreover, all blocks in a stack are not reducers except the
last one. As such, each stack can increase or decrease the number of
channels in the data representation and at the same time it decreases the size
of the data with only one convolution. Figure~\ref{fig:ch_formal_nn_model}
represents a stack.

\begin{figure}
	\vspace*{0mm}
	\center
	\includegraphics[width=0.35\textwidth]{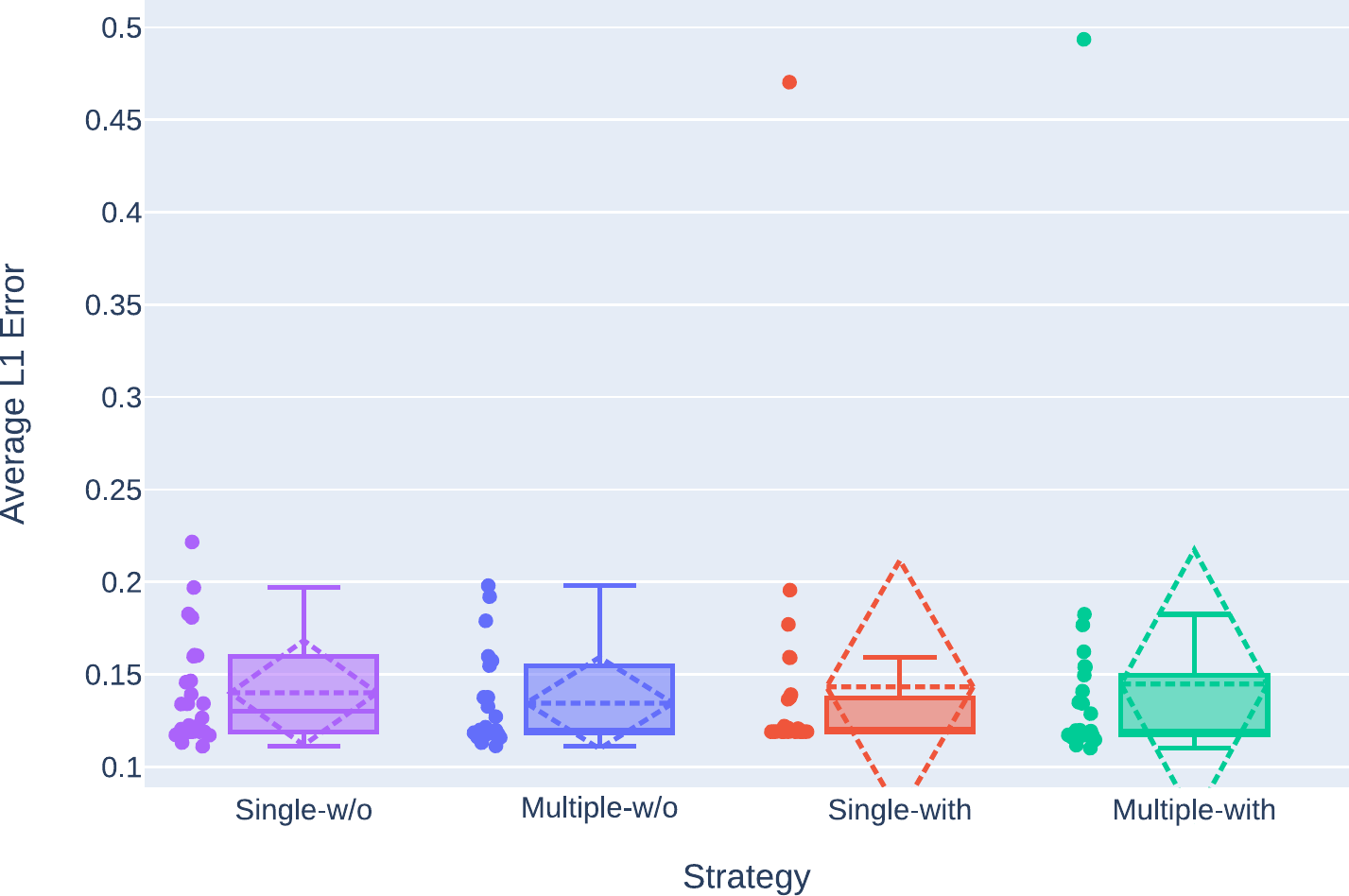}
	\caption{
		Comparison between the 4 proposed strategies. \textbf{Single-w/o} strategy
		corresponds to the traditional non acoustics-specific transcription function. The
		dotted rhombus represents average and standard deviation. The continuous
		line represents the median and box-plot.
	}
	\label{fig:ch_formal_vel_results_1}
\end{figure}

In order to control the complexity of the network, we designed a family of CNNs
that vary the ratio between the number of blocks and the number of channels in
each stack based on two parameters $k_1$ and $k_2$. By setting $k_1$ and $k_2$,
one may find Residual CNNs that approximate various types of functions.
Specifically, $k_1$ is inversely proportional to $k_2$ while $k_1$ is related to
the number of blocks and $k_2$ is connected with the number of channels -- see
Fig.~\ref{fig:ch_formal_nn_model}.
% the algorithm used for shaping one CNN operates as follows.  In each stack, the number
% of blocks is defined as $max(1, \lfloor \frac{2^{k_1}}{l} \rceil)$, where the
% $\lfloor \cdot \rceil$ represents the rounding operation, $l$ is the number of
% input channels and $k_1$ is an hyper-parameter. Similarly, the number of output
% channels in each stack was computed as $l \times \lfloor k_2 \rceil$. For
% instance, the first stack will always have $\lfloor 2^{k_1} \rceil$ blocks and
% $\lfloor k_2 \rceil$ output channels, because the input channel size is 1; the
% $i$-th stack, instead, will have $\lfloor \frac{2^{k_1}}{\lfloor k_2
% 		\rceil^{(i-1)}} \rceil$ blocks and $\lfloor k_2 \rceil^i$ channels. 
In our experiments, we manually found that $k_1 = 4$ comprises an effective
parameter and observed the way the models perform when $k_2$ changes.  Following
this algorithm, multiple stacks were concatenated until the output size has at
least one dimension $< k_0$, where $k_0$ is the kernel size, which is fixed
across the stacks.

We use multiple of such CNNs in each model to estimate encoder and performer
functions and an additional one for the context classifier.  After the stacks, a
further convolutional layer followed by batch normalization and ReLU is added
aiming at reducing the data size to 1 and at compressing all existing
information into the channel dimension; in the performer and context classifier,
this last block also takes care of reducing the number of channels to the
expected output dimension, i.e. 1 for the velocity and 6 for the context
classifier.  Finally, we apply a linear transformation using a grouped
convolution and activation block with kernel size 1; the last activation is a
ReLU in the encoder, a Sigmoid in the performers, and a SoftMax in the context
classifier.

% \begin{figure*}
% 	\center
% 	\includegraphics[width=0.75\textwidth]{strategies.pdf}
% 	\caption{
% 		The 4 strategies tested in this work. Each strategy corresponds to a
% 		different model architecture during training. See
% 		Figure~\ref{fig:ch_formal_nn_model} for the representation of a
% 		\textit{ConvStack}. $C_o$ is the number of outputs of each \textit{ConvStack}.
% 	}
% 	\label{fig:ch_formal_strategies}
% \end{figure*}

The considered hyper-parameters were 4:

\begin{enumerate}

	\item the kernel size in the encoder (values: {3, 5})
	\item the kernel size in the performer (values: {3, 5})
	\item the $k_2$ parameter in the encoder (values: {1, 2, 3})
	\item the $k_2$ parameter in the performer (values: {1, 2, 4})

\end{enumerate}

The acoustic context classifier branch is built with the same performer kernel; however,
due to the higher computational complexity needed for classifying 6 labels, $\{k_1,~
	k_2\}$ were multiplied by 1.25 -- i.e. $k_1 = 5$ and $\lfloor k_2 \rceil \in
	\{1, 3, 5\}$.

\section{Training}

The training datasets include 7.1 millions of music notes. Since the experiment
includes the training of more than 100 models, to make the problem
computationally accessible, we use only 0.1\% of the available data with a batch
size of 10, resulting in 703 batches (7030 notes). Subsampling was performed
with a uniform distribution and was repeated on all 6 contexts and splits
(train, validation, and test sets). Overall, the training set is made of 566
batches, the validation counted 63 batches, and the test set is composed of 74
batches.

Training is performed using Adadelta~\cite{zeiler2012adadelta} optimizer
with initial learning rate set according to an existing algorithm designed to
find its optimal value based on repeated small experiments with increasing
learning rates~\cite{smith2017cyclical}. When the algorithm fails, the initial
learning rate is automatically set to 1e-5. The loss function for the
performers is the L1 error, while for the context classifier we use the
Cross-Entropy loss. When the context classifier is used, we
treat the problem from a multi-task perspective. For this reason, we sum
the two losses and use the recently proposed RotoGrad
algorithm~\cite{javaloy2021rotograd} to
stabilize the gradients. Moreover, when using multiple performers, to speed up the
training process, we load data so that each batch contains notes related to only one
context at a time and we cycle across contexts so that all of them are equally
represented. However, this strategy leads to unstable losses both in training
and in validation, making it hard to understand when the model is actually
overfitting. As such, we imposed an early-stop procedure with a patience
of 20 by observing the Exponential Moving Average of the validation loss on a
window of 15 epochs.

\section{Results}
\label{sec:ch_formal_results}

\begin{figure}
	\center
	\includegraphics[width=0.35\textwidth]{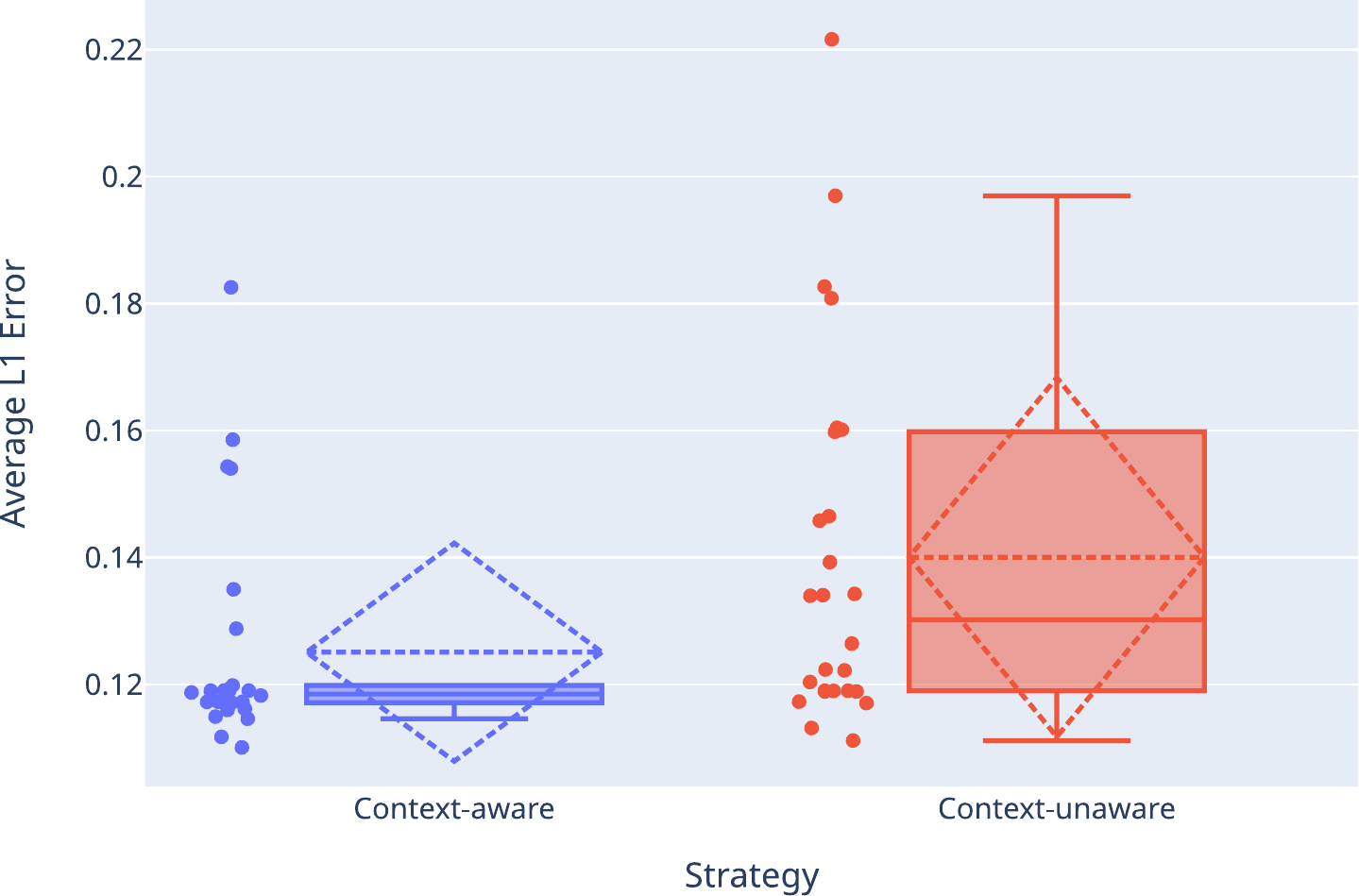}
	\caption{
		Comparison between the best acoustics-specific strategy and the traditional
		non acoustics-specific transcription in each hyper-parameter point. The dotted
		rhombus represents average and standard deviation. The continuous line
		represents the median and box-plot.
	}
	\label{fig:ch_formal_vel_results_2}
\end{figure}

The results we obtained are shown in
Figure~\ref{fig:ch_formal_vel_results_1}~and~\ref{fig:ch_formal_vel_results_2}.
We computed the average L1 error for velocity estimation in each tested
hyper-parameter set, discarding those configurations that generated models
exceeding GPU or even CPU RAM or that returned invalid losses. Overall, we
considered 26 hyper-parameter sets corresponding to 104 runs. Moreover,
to reduce the computational burden, we stopped each training at the 40th epoch,
in case the training procedure was not terminated by the early-stopping criterion.

\begin{table}
	\caption{Win analysis. The table must be read as: ``strategy at row \textit{x} is better
		than strategy at column \textit{y} in \textit{n} hyper-parameter
		configurations''}
	\begin{center}
		\scriptsize
		\begin{tabular}{|c|ccccc|}
			\hline
			                       & \textbf{Single-w/o}     & \textbf{Multiple-w/o}   & \textbf{Single-with}    & \textbf{Multiple-with}  & \textbf{All} \\ \hline
			\textbf{Single-w/o}    & \multicolumn{1}{c|}{-}  & \multicolumn{1}{c|}{2}  & \multicolumn{1}{c|}{12} & \multicolumn{1}{c|}{3}  & 1            \\ \cline{2-6}
			\textbf{Multiple-w/o}  & \multicolumn{1}{c|}{24} & \multicolumn{1}{c|}{-}  & \multicolumn{1}{c|}{19} & \multicolumn{1}{c|}{11} & 11           \\ \cline{2-6}
			\textbf{Single-with}   & \multicolumn{1}{c|}{14} & \multicolumn{1}{c|}{7}  & \multicolumn{1}{c|}{-}  & \multicolumn{1}{c|}{6}  & 5            \\ \cline{2-6}
			\textbf{Multiple-with} & \multicolumn{1}{c|}{23} & \multicolumn{1}{c|}{12} & \multicolumn{1}{c|}{20} & \multicolumn{1}{c|}{-}  & 12           \\ \hline
		\end{tabular}
	\end{center}
	\label{tab:ch_formal_velocity_wins}
\end{table}

For evaluating the statistical significance of the results in
Figure~\ref{fig:ch_formal_vel_results_1}, we applied the Shapiro-Wilk normality
test to each strategy distribution and then Kruskal-Wallis and Wilcoxon
signed-rank test for post-hoc analysis. We found that all  analyzed
distributions rejected the null hypothesis of normality tests with $p <
	4\text{e-}3$, meaning that the distributions are not normal. We found no
significant difference according to the omnibus Kruskal-Wallis test ($p =
	1.74\text{e-}1$). However, we also computed the Wilcoxon p-values using the
Bonferroni-Holm correction and found a statistically significant difference with
confidence of 95\% only between Multiple-w/o and Single-w/o, Multiple-with and
Single-w/o. Note that the $p > 0.05$ found with the Kruskal-Wallis test is
coherent with the pairwise significance found using the corrected Wilcoxon
test~\cite{hsu1996multiple,maxwell2017designing}. Given the statistical
analysis, we  argue that $R > 0$ for the Multiple-with and Multiple-w/o strategies. To
further assess such conclusions, we also computed the number of hyper-parameter
sets won by each strategy. Table~\ref{tab:ch_formal_velocity_wins} shows this
analysis and highlights how in only 1 configuration the best training strategy
was Single-w/o.

However, no optimal strategy was found. Indeed, considering Multiple-w/o,
Single-with, and Multiple-with, there is no agreement about which one is the most
effective method. To obtain a deeper understanding of the problem, we tried to
check what would happen in case an oracle could indicate the optimal strategy
depending on the model shape. Results were highly statistically significant ($p
	= 6\text{e-}5$) and showed far improved results when one of the proposed
acoustics-specific strategy was used -- see Figure~\ref{fig:ch_formal_vel_results_2}.
This result highlights how, in theory, the benefit coming from acoustics-specific AMT
can be definitely larger than 0.

\section{Conclusions}
\label{sec:ch_formal_conclusions}

In this paper, we proposed an extensible framework to deepen the understanding
of acoustic factors on music performance analysis from the perspective of AMT.

We extensively evaluated 4 different strategies for velocity
estimation of single notes. We demonstrated that considering acoustics-specific
strategies consistently improves model performance. However, no acoustics-specific
strategy was found to outperform the rest; it was shown that they complement
each other.

Future works could also estimate non-MIDI parameters that are relevant for the
timbre realization of pianists~\cite{bernays2014investigating}. Another
attractive addition would be the note offset precise inference based on the
hammer second and third impulsive sound; given the low accuracy of the note offset
inference in state-of-the-art AMT models, such an addition could be useful for
precisely defining the performer interpretation. A third addition could be
performer-specific adaptation functions, as suggested in previous
experiments~\cite{kalkandjiev2015influence}. Finally, an important parameter
that we plan to focus in next works is the pedaling level estimation.

\balance
\bibliographystyle{jabbrv_ieeetr}
\bibliography{IEEEabrv.bib,bibliography.bib}

\end{document}